\documentclass{wscpaperproc}
\usepackage{latexsym}
\usepackage{graphicx}
\usepackage{mathptmx}
\usepackage[T1]{fontenc}

\usepackage{algorithm}
\usepackage{algorithmic}
\usepackage{amsmath}

\usepackage{amsmath}
\usepackage{amsfonts}
\usepackage{amssymb}
\usepackage{amsbsy}
\usepackage{amsthm}
\usepackage{xcolor} % For coloring code
\usepackage{subfigure}
\usepackage{graphicx}

\usepackage{listings}
\usepackage{xcolor}  % For color highlighting

\lstdefinelanguage{json}{
    basicstyle=\ttfamily\small,
    numbers=left,
    numberstyle=\tiny,
    stepnumber=1,
    showstringspaces=false,
    breaklines=true,
    frame=lines,
    backgroundcolor=\color{gray!10},
    morestring=[b]",
    morecomment=[l]{//},
    moredelim=[l][\color{blue}]{:}
}

\newcommand{\note}[1]{{\color{black}{#1}}}

\lstdefinestyle{mystyle2}{
    backgroundcolor=\color{white},   
    commentstyle=\color{gray},
    keywordstyle=\color{blue},
    numberstyle=\tiny\color{gray},
    stringstyle=\color{black},
    basicstyle=\ttfamily\footnotesize,
    breakatwhitespace=false,         
    breaklines=true,                 
    captionpos=b,                    
    keepspaces=true,                 
    numbers=left,                    
    numbersep=5pt,                  
    showspaces=false,                
    showstringspaces=false,
    showtabs=false,                  
    tabsize=2
}

\usepackage[pdftex,colorlinks=true,urlcolor=blue,citecolor=black,anchorcolor=black,linkcolor=black]{hyperref}

\newtheoremstyle{wsc}% hnamei
{3pt}% hSpace abovei
{3pt}% hSpace belowi
{}% hBody fonti
{}% hIndent amounti1
{\bf}% hTheorem head fontbf
{}% hPunctuation after theorem headi
{.5em}% hSpace after theorem headi2
{}% hTheorem head spec (can be left empty, meaning `normal')i

\theoremstyle{wsc}

\begin{document}

%***************************************************************************
% AUTHOR: AUTHOR NAMES GO HERE
% FORMAT AUTHORS NAMES Like: Author1, Author2 and Author3 (last names)
%
%		You need to change the author listing below!
%               Please list ALL authors using last name only, separate by a comma except
%               for the last author, separate with "and"
%

% setting up general page style
\pagestyle{fancyplain}

% setting up page style of first page
\thispagestyle{plain}
\firstPageHead{}

% setting up running header (authors) of subsequent pages
\chead{\fancyplain{}{\itshape Abdurahman, Hossain, Brown, Yoshii, and Ahmed}}

% setting up seperation parameters
%\headsep=72pt
\rhead{}
\cfoot{}
\renewcommand{\headrulewidth}{0pt} % (renewcommand needed in fancyhdr to remove top decorative line)
%\headrulewidth=0pt  % ("setlength" needed in fancyheading to remove top decorative line)

%%%%%%%%%%%%%%%%%%%%%%%%%%%%%%%%%%%%%%%%%%%%%%%%%%%%%%%%%%%%%%%%%%%%%%%%%%%%%%
%                                                                            %
%     THESE COMMANDS ARE REQUIRED TO WORK WITH WSC.BST TO MAKE BIBLIO     %
%                                                                            %
%%%%%%%%%%%%%%%%%%%%%%%%%%%%%%%%%%%%%%%%%%%%%%%%%%%%%%%%%%%%%%%%%%%%%%%%%%%%%%
\makeatletter
\let\@internalcite\cite
\def\cite{\def\@citeseppen{-1000}%
    \def\@cite##1##2{(##1\if@tempswa , ##2\fi)}%
    \def\citeauthoryear##1##2##3{##1 ##3}\@internalcite}
\def\citeNP{\def\@citeseppen{-1000}%
    \def\@cite##1##2{##1\if@tempswa , ##2\fi}%
    \def\citeauthoryear##1##2##3{##1 ##3}\@internalcite}
\def\citeN{\def\@citeseppen{-1000}%
%  Pierre L'Ecuyer's fix for multiple cite bug
%  Added by Paul J Sanchez on 4 October 2001
%   \def\@cite##1##2{##1\if@tempswa , ##2)\else{)}\fi}%
%   \def\citeauthoryear##1##2##3{##1 (##3}\@citedata}
    \def\@cite##1##2{##1\if@tempswa, ##2)\else{}\fi}%
    \def\citeauthoryear##1##2##3{##1 (##3)}\@citedata}
\def\citeA{\def\@citeseppen{-1000}%
    \def\@cite##1##2{(##1\if@tempswa , ##2\fi)}%
    \def\citeauthoryear##1##2##3{##1}\@internalcite}
\def\citeANP{\def\@citeseppen{-1000}%
    \def\@cite##1##2{##1\if@tempswa , ##2\fi}%
    \def\citeauthoryear##1##2##3{##1}\@internalcite}
\def\shortcite{\def\@citeseppen{-1000}%
    \def\@cite##1##2{(##1\if@tempswa , ##2\fi)}%
    \def\citeauthoryear##1##2##3{##2 ##3}\@internalcite}
\def\shortciteNP{\def\@citeseppen{-1000}%
    \def\@cite##1##2{##1\if@tempswa , ##2\fi}%
    \def\citeauthoryear##1##2##3{##2 ##3}\@internalcite}
\def\shortciteN{\def\@citeseppen{-1000}%
%  Pierre L'Ecuyer's fix for multiple cite bug
%  Added by Paul J Sanchez on 2 September 2002
%  should have caught this last year...
%   \def\@cite##1##2{##1\if@tempswa , ##2)\else{)}\fi}%
%   \def\citeauthoryear##1##2##3{##2 (##3}\@citedata}
% Shane G. Henderson fix for extra right bracket at end of optional material June 8, 2005
%    \def\@cite##1##2{##1\if@tempswa, ##2)\else{}\fi}%
    \def\@cite##1##2{##1\if@tempswa, ##2\else{}\fi}%
    \def\citeauthoryear##1##2##3{##2 (##3)}\@citedata}
\def\shortciteA{\def\@citeseppen{-1000}%
    \def\@cite##1##2{(##1\if@tempswa , ##2\fi)}%
    \def\citeauthoryear##1##2##3{##2}\@internalcite}
\def\shortciteANP{\def\@citeseppen{-1000}%
    \def\@cite##1##2{##1\if@tempswa , ##2\fi}%
    \def\citeauthoryear##1##2##3{##2}\@internalcite}
\def\citeyear{\def\@citeseppen{-1000}%
    \def\@cite##1##2{(##1\if@tempswa , ##2\fi)}%
    \def\citeauthoryear##1##2##3{##3}\@citedata}
\def\citeyearNP{\def\@citeseppen{-1000}%
    \def\@cite##1##2{##1\if@tempswa , ##2\fi}%
    \def\citeauthoryear##1##2##3{##3}\@citedata}
%
% \@citedata and \@citedatax:
%
% Place commas in-between citations in the same \citeyear, \citeyearNP,
% \citeN, or \shortciteN command.
% Use something like \citeN{ref1,ref2,ref3} and \citeN{ref4} for a list.
%
\def\@citedata{%
    \@ifnextchar [{\@tempswatrue\@citedatax}%
                  {\@tempswafalse\@citedatax[]}%
}

\def\@citedatax[#1]#2{%
\if@filesw\immediate\write\@auxout{\string\citation{#2}}\fi%
  \def\@citea{}\@cite{\@for\@citeb:=#2\do%
    {\@citea\def\@citea{, }\@ifundefined% by Young
       {b@\@citeb}{{\bf ?}%
       \@warning{Citation `\@citeb' on page \thepage \space undefined}}%
{\csname b@\@citeb\endcsname}}}{#1}}%

% don't box citations, separate with ; and a space
% also, make the penalty between citations negative: a good place to break.
%
\def\@citex[#1]#2{%
\if@filesw\immediate\write\@auxout{\string\citation{#2}}\fi%
  \def\@citea{}\@cite{\@for\@citeb:=#2\do%
    {\@citea\def\@citea{; }\@ifundefined% by Young
       {b@\@citeb}{{\bf ?}%
       \@warning{Citation `\@citeb' on page \thepage \space undefined}}%
{\csname b@\@citeb\endcsname}}}{#1}}%

% (from apalike.sty)
% No labels in the bibliography.
%
\def\@biblabel#1{}
\makeatother

%\newlength{\bibhang}
%\setlength{\bibhang}{2em}

% Indent second and subsequent lines of bibliographic entries. Taken
% from openbib.sty: \newblock is set to {}.
% \renewcommand{\refname}{REFERENCES}

\newdimen\bibindent
\bibindent=0.0em
% SEC: was \def\thebibliography#1{\section*{\refname\@mkboth
% SEC: was   {\uppercase{\refname}}{\uppercase{\refname}}}\list
\def\thebibliography#1{\section*{\refname}\list
   {}{\settowidth\labelwidth{[#1]}
   \leftmargin\parindent
   \itemindent -\parindent
   \listparindent \itemindent
   \itemsep 0pt
   \parsep 0pt}
   \def\newblock{}
   \sloppy
   \sfcode`\.=1000\relax}

           % Set up BiBTeX macros

% needed to make the tex document look more like the word counterpart :-(
\setlength{\baselineskip}{12.7pt}

% AUTHOR: Enter the title, all letters in upper case
\title{Scalable HPC Job Scheduling and Resource Management in SST}

% AUTHOR: Enter the authors of the article, see end of the example document for further examples
\author{
    Abubeker Abdurahman\textsuperscript{1}, Abrar Hossain\textsuperscript{1}, Kevin A. Brown\textsuperscript{2},  
    Kazutomo Yoshii\textsuperscript{2}, and Kishwar Ahmed\textsuperscript{1} \\[10pt]
    \textsuperscript{1}Department of Electrical Engineering and Computer Science, The University of Toledo, Toledo, OH, USA \\ 
    \textsuperscript{2}Mathematics and Computer Science Division, Argonne National Laboratory, Lemont, IL, USA
}

\maketitle

\vspace{-12pt}  % If necessary, keep it but ensure it does not break formatting.

\begin{abstract}
% Update the abstract with workflow stuff
  Efficient job scheduling and resource management contributes towards system throughput and efficiency maximization in high-performance computing (HPC) systems. In this paper, we introduce a scalable job scheduling and resource management component within the structural simulation toolkit (SST), a cycle-accurate and parallel discrete-event simulator. 
  % Central to our approach is the integration of a TaskScheduler with a dynamic ResourceManagement component, enabling a detailed simulation of CPU \& memory resource distribution \& task scheduling across computing nodes.
  % Our proposed simulator can be used for comprehensive analysis of diverse HPC workload. 
  Our proposed simulator includes state-of-the-art job scheduling algorithms and resource management techniques. Additionally, it introduces a workflow management components that supports the simulation of task dependencies and resource allocations, crucial for workflows typical in scientific computing and data-intensive applications. 
  % We incorporate state-of-the-art different job scheduling algorithms in our simulator. 
  % The simulator can be used to perform what-if analysis of HPC job scheduling \& resource management techniques. 
  % The proposed simulator is incorporated within the SST components, \& can be used for realistic application performance behavior analysis with the help of SST simulation. The workflow-based scheduling within SST facilitates modeling of HPC application with workflow dependencies.  
  % The integration of workflow-based scheduling within SST extends its applicability, making it a powerful tool for modeling the performance of complex HPC systems. 
  We present validation and scalability results of our job scheduling simulator. Simulation shows that our simulator achieves good accuracy in various metrics (e.g., job wait times, number of nodes usage) and also achieves good parallel performance.  
  % We uncover the impact of varying task scheduling \& resource management strategies on pivotal performance metrics, including CPU \& memory utilization, job wait times, \& overall system throughput. 
  % The innovation of our research lies in its unique combination of scheduling algorithms \& a resource management framework, offering a scalable \& adaptable solution to HPC challenges..
  \end{abstract}

\section{INTRODUCTION}
\label{sec:intro}
% Update the intro with 
High-performance computing (HPC) systems have witnessed unprecedented growth in recent years, not only in the sheer system size but also in their architectural complexity and their sophisticated HPC application processing capabilities. 
Many advanced job scheduling and resource management techniques have been developed to achieve desired performance objectives of complex scientific applications running on large HPC systems. 
These techniques aim to increase both resource utilization and job turnaround times, thereby justifying the increasing resource demands, ranging from compute nodes and memory to power, of next-generation HPC systems. 
With increase in larger and computationally-demanding applications (e.g., machine learning workloads), it is important that we have efficient scheduling and resource management evaluation techniques in HPC~\shortcite{soysal2021collection}. %~\cite{Sanchez2011AgentBased}  

Simulation emerges as an effective tool in evaluating the efficiency of job scheduling and resource management techniques as it offers insights into these techniques' performance implications on overall HPC system's performance and utilization. A number of job scheduling simulators exist. For example, CQsim~\cite{CQsim}, Alea 2~\cite{Klusacek2010Alea2}, Slurm Simulator~\shortcite{simakov2018slurm}, and GridSim~\cite{Buyya2002Gridsim} are some of the existing job scheduling simulators. CQsim excels in event-driven job scheduling with its diverse input options. Alea 2 supports most of the common scheduling algorithms and includes an intuitive visualization interface. The Slurm simulator enhances modeling fidelity by mirroring the Slurm resource manager's~\shortcite{yoo2003slurm} functionality without taxing the actual system performance. GridSim's Java-based framework adeptly models heterogeneous grid environments, focusing on the distributed management of resources. However, despite their respective strengths, common shortfalls among these simulators are their limited scalability and the absence of comprehensive scheduling algorithms essential for simulating the dynamic HPC job behavior. We propose an HPC job scheduling simulator that integrates state-of-the-art HPC job scheduling algorithms, and achieves scalability and high fidelity on simulating HPC jobs, resources, and workflows.
% ~\cite{Zheng2004BigSim, Law2014SimulationModelingAnalysis}

The structural simulation toolkit (SST) is a well-known architecture and network simulator that supports both cycle-accurate and parallel discrete-event based simulation~\shortcite{rodrigues2011structural}. SST's scalability is underscored by its modular design. SST contains three libraries -- \texttt{SST Core}, \texttt{SST Elements}, and \texttt{SST Macros}. The \texttt{SST Core} facilitates the fundamental simulation framework. The \texttt{SST Elements} provides the building blocks for modeling complex systems. The \texttt{SST Macros} facilitates scalable and large-scale system simulations.
% SST's architecture includes specialized modules such as \texttt{memHierarchy} for memory simulation \& \texttt{Merlin} for network modeling -- offering detailed insights into system interactions~\cite{sst-elements}. These features enable comprehensive simulations of memory \& networking within HPC environments, enhancing the realism of modelled scenarios. 
SST offers parallel execution and scalability by employing conservative synchronization strategy~\shortcite{rodrigues2011structural}.
% SST employs conservative synchronization, ensuring accurate \& efficient simulation across multiple computational nodes, highlighting its capacity for parallel execution \& scalability.

In this paper, we introduce a novel job scheduling and resource management component within SST. 
% addressing gaps in task scheduling \& resource management previously unsupported by the framework. 
% Our simulator also incorporates a resource management component. 
It supports five state-of-the-art job scheduling algorithms: first come first serve (\texttt{FCFS}), shortest job first (\texttt{SJF}), longest job first (\texttt{LJF}), \texttt{FCFS} with \texttt{Best Fit} and \texttt{FCFS} with \texttt{Backfilling}. 
% This suite of algorithms enhances SST's utility in modeling diverse HPC scheduling scenarios, a feature absent in prior SST versions.
% ~\cite{Smith1978}
% Our Job Scheduler's uniqueness lies not only in the breadth of supported algorithms but also in its 
Our simulator is integrated within SST's discrete event-driven simulation engine. Each job goes through detailed simulation of job lifecycle management (e.g., submission, execution, completion) across various system configurations. The simulation environment can be configured using SST parameters to accurately represent diverse HPC architectures and resource availabilities. 
% These parameters include the number of nodes, CPU cores per node, \& memory capacity, enabling flexible adjustments to simulate different network environments \& resource configurations effectively. 
\note{Compared to other simulators, such as CQsim and Alea 2, our simulation enhances scalability and supports a comprehensive range of job scheduling algorithms. Furthermore, it leverages SST’s unique capabilities for high fidelity simulations.}
\note{Furthermore, the existing job scheduling simulator cannot support workflow management of complex and interdependent workflow-based science applications.}

% its capacity to improve simulation performance. Our results confirm the scheduler's scalability \& its potential to significantly enhance SST's capabilities ~\cite{GWA_DAS2_2023}.

We extend the capabilities of SST by introducing a workflow management component tailored for representing task dependencies typical in scientific computing. This addition complements our existing job scheduling and resource management components by enabling the detailed simulation of workflows. 
% that are characteristic of real-world applications in HPC environments. 
The workflow management feature within SST supports basic task scheduling algorithms (e.g., \texttt{FCFS}) and captures essential workflow dynamics through the discrete event-driven simulation engine. Workflow tasks are modeled to respect interdependencies, allowing the system to handle submission, execution, and completion phases dynamically. 
We configure the simulation environment with SST parameters that accurately mimic real-world computing architectures.
% , enhancing the toolkit's applicability for performance analysis. 

We validated our proposed simulator against CQsim (a cluster scheduling simulator) using traces collected from Grid Workload Archive and Parallel Workload Archive. We also performed performance testing of our simulator. The simulation results show that our simulator is accurate in job scheduling and is able to achieve good parallel performance. 
We also conducted  validation of the workflow component using simplified workflow models based on well-known workflow analysis system such as Pegasus~\shortcite{deelman2015pegasus} and Apache Airflow~\cite{PegasusGallery2024}. 
% We aim to enhance these features by integrating advanced scheduling capabilities \& dynamic resource allocation strategies, thereby improving SST’s utility for simulating \& managing complex \& detailed workflows in HPC settings. 
Simulation indicates that the workflow component adheres effectively to task dependency constraints, while maintaining good performance across simulations.

% \note{In the development of our job scheduling simulator within the SST framework, we integrate both traditional and advanced scheduling algorithms such as first come first serve, and Backfilling with resource allocation procedures that adapt to the simulated system's state. This integration is a sophisticated enhancement that leverages SST’s unique capabilities for high fidelity simulations. Our simulator offers a more granular control over resource allocation \& task prioritization, tailored specifically for high-performance computing environments.} 
% \note{Compared to other simulators such as CQsim and Alea 2, it enhances scalability and supports a comprehensive range of job scheduling algorithms.} 
% \note{The workflow support within SST allows a wide variety of simulations, from simple batch processing to complex and interdependent scientific workflows, and further differentiates from existing job scheduling simulators.}
% \note{Built on the modular and highly extensible SST platform, our tool is uniquely capable of accommodating a broad spectrum of HPC simulations, thus providing significant advancements in handling dynamic HPC job behaviors.}

\section{Simulator Design}

\begin{figure}[t!]
    \centering
    \includegraphics[width=0.75\textwidth]{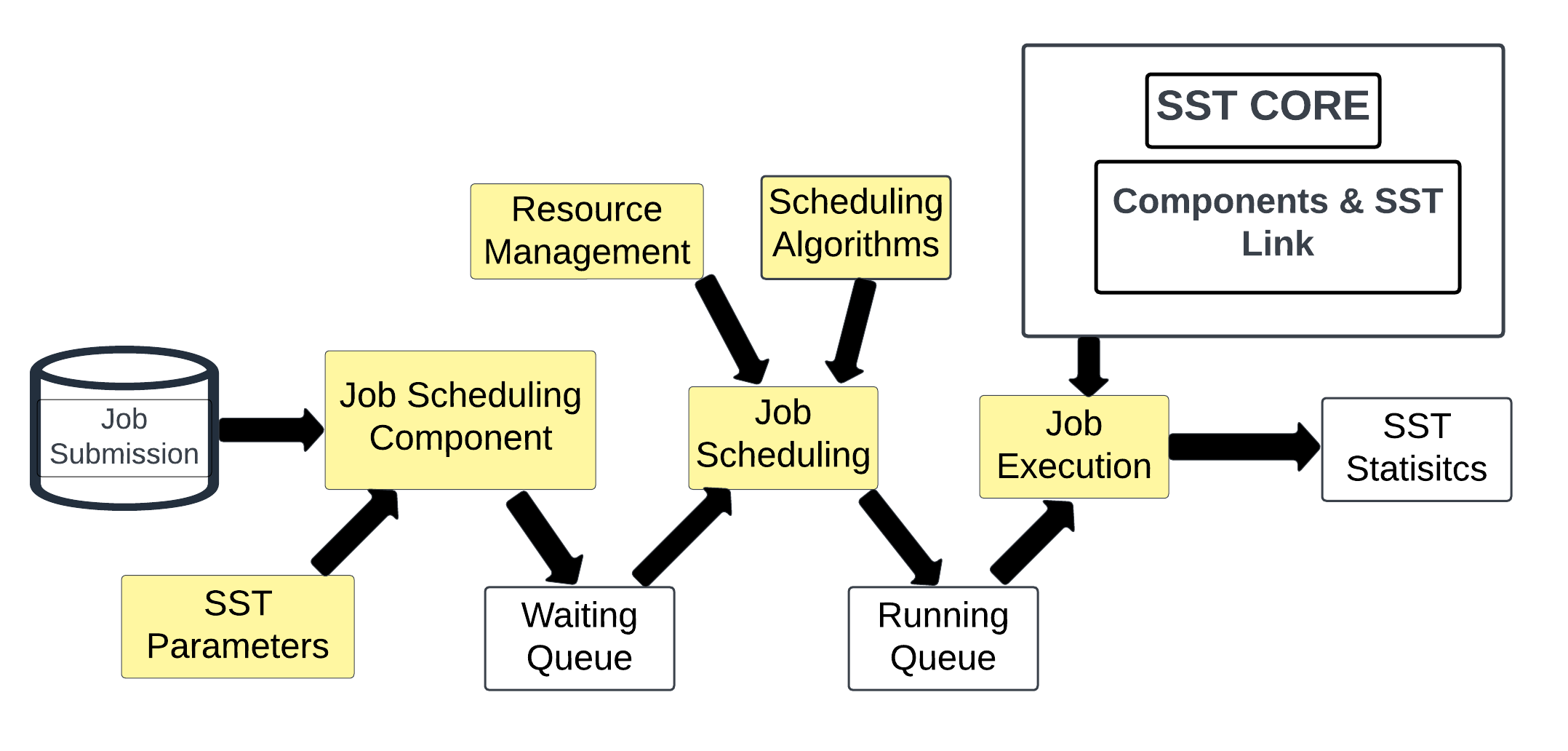}
    \caption{The simulator components.}
    \label{fig:SST_workflow}
\end{figure}

The overall simulator diagram is shown in Figure~\ref{fig:SST_workflow}. 
The \texttt{SST Core} initializes the simulation, creating the necessary components and establishing links. 
The \texttt{Job Scheduling} module receives information about jobs from the components and places them in the queue based on their priorities and dependencies.
% Jobs will be assigned from the event queue. \texttt{Job Scheduler} manages the distribution based on the current \texttt{Resource Pool} status and the simulation's Clock.
% As tasks are executed, the \texttt{Job Scheduler} receives feedback on resource utilization and job completion times.
After a job is submitted, it will enter the job waiting queue and \texttt{Job Scheduling} module will be invoked. The \texttt{job Scheduling} module will determine which jobs to run from the waiting queue based on the scheduling policies and the available resources from the \texttt{Resource Management} module. When a job is scheduled to run the \texttt{job Scheduling} will remove the job from the waiting queue and will put it in the running queue. The \texttt{Job Executor} will allocate the resources using the \texttt{Resource Management}. The \texttt{Job Executor} will then simulate the job execution. When a job completes its execution, the \texttt{Job Executor} will remove the job from the list of running jobs and will reclaim the resources occupied by the job. We describe each component in detail next.
% We describe the details of each simulator component next.

\subsection{Job Arrival and Scheduling}

Upon arrival, each job is encapsulated as a \texttt{TaskEvent} class instance. Unique job identifiers and detailed resource requirements are assigned to each instance. 
% This encapsulation allows for nuanced scheduling of jobs, tailored to the simulation's real-time resource landscape. 
The \texttt{Job Scheduling} module, in coordination with the \texttt{Resource Management} module assesses available and required resources, and ensures job queueing and scheduling. 
%\note{Abubeker: please put the job attributes (job id, job size) and related description}.
% This process not only streamlines job submission but also sets the stage for a flexible scheduling mechanism capable of adapting to varying computational demands.
 Serialization within the \texttt{TaskEvent} class ensures accurate transfer of task data across SST components. We show the \texttt{TaskEvent} class below.

\begin{lstlisting}[language=C++, caption={Serialization method implementation for TaskEvent class.}, label=lst:task-event-serialization]
TaskEvent(const std::string& id, int cores, double time, size_t memory) : 
        Event(), jobID(id), requiredCores(cores), executionTime(time), requiredMemory(memory), state(QUEUED) {}

    void serialize_order(SST::Core::Serialization::serializer& ser) override {
        Event::serialize_order(ser);
        ser & jobID;
        ser & requiredCores;
        ser & executionTime;
        ser & requiredMemory;
        ser & state; 
    }
    ImplementClassName("TaskEvent");
\end{lstlisting}

% \subsection{Job Scheduling}

Our simulator includes a range of job scheduling algorithms to support state-of-the-art HPC job scheduling techniques. 
We support five job scheduling algorithms: \texttt{FCFS}, \texttt{FCFS} with \texttt{Backfilling}, \texttt{FCFS} with \texttt{Best Fit,} \texttt{SJF}, and \texttt{LJF}. Each job scheduling algorithm satisfies diverse objectives (e.g., minimizing wait times, optimizing resource utilization), and reflects the multifaceted nature of job scheduling in HPC environments. We outline each job scheduling algorithm details below.
    \begin{itemize}

        \item \textbf{First-Come, First-Served (\texttt{FCFS}):} This technique schedules jobs in the order they arrive. This straightforward and fair approach ensures that jobs are processed in a sequence that they arrive, however, it may not always lead to the most efficient use of system resources, especially in systems with diverse job duration.
            
        \item \textbf{\texttt{FCFS} with \texttt{Backfilling}:} This  algorithm enhances job throughput by allowing smaller jobs to proceed when larger jobs are waiting for resource availability. This method optimizes the overall utilization of system resources by dynamically adjusting the queue based on job size and available resources.
        
        \item \textbf{\texttt{FCFS} with \texttt{Best} \texttt{Fit}:} This technique allocates resources based on the closest match to the job’s requirements, minimizing wastage and improving utilization. This strategy is particularly effective in environments where resource heterogeneity can lead to inefficiencies in allocation and utilization.
        
        \item \textbf{Longest Job First (\texttt{LJF}):} This algorithm prioritizes longer tasks to potentially optimize certain types of workloads or system conditions. The approach is less common. However, it can be beneficial in scenarios where longer jobs need to be expedited, albeit at the risk of increasing waiting times for shorter jobs.
        
        \item \textbf{Shortest Job First (\texttt{SJF}):} This technique minimizes average wait time by prioritizing shorter tasks. \texttt{SJF} is effective in homogeneous environments but hinges on the accurate estimation of job lengths, which can be challenging to ascertain~\cite{Smith1978}.
    \end{itemize}

There are more recent and advanced scheduling techniques that can improve the job scheduling performance. For example, the deep reinforcement agent for scheduling (DRAS) technique proposed in~\shortcite{fan2022dras} improves the job's adaptability and performance further. Although not currently implemented, we plan to implement such techniques for job scheduling techniques with better performance. 
% In our pursuit of more advanced and efficient scheduling solutions, we are exploring the integration of novel algorithms, including Deep Reinforcement Agent for Scheduling (DRAS). These cutting-edge strategies, highlighted in recent studies ~\cite{fan2022dras}, promise to enhance our scheduler's adaptability and performance further. Our ongoing efforts aim to expand our scheduler's capabilities, incorporating Reinforcement Learning (RL) based scheduling to dynamically adjust to the simulation environment, ensuring optimal decision-making and resource allocation ~\cite{fan2022dras}.

\subsection{Job Execution and Resource Allocation}

Upon determining the execution order, the \texttt{Job Scheduling} module allocates resources to jobs using the selected scheduling algorithm. The \texttt{Job Executor} module executes the jobs using SST. 
Algorithm~\ref{alg:resource-allocation-deallocation} outlines how our simulator determines resources and executes jobs within an HPC environment. Job events are prioritized and managed in a dynamic priority queue. The simulator initializes predefined CPU resources and handles job submissions by enqueuing them in the priority queue. It allocates the necessary cores and schedules them for execution. As jobs are completed, resources are reclaimed and reallocated to pending jobs in the queue. We can evaluate the efficacy of various scheduling strategies in an HPC context with the help of continuous allocation and deallocation of job resources.

% \begin{lstlisting}[label=lst:Resource allocation/deallocation]
% Initialize:
% Define totalCores, availableCores
% - Create an empty priority queue for JobEvents based on their priority

% Function scheduleJobEvent(jobEvent):
% - Add jobEvent to the priority queue
% - Attempt to allocate resources for jobEvent

% Function allocateResources(jobEvent):
% - If availableCores >= jobEvent.requiredCores:
% - Deduct jobEvent.requiredCores from availableCores
% - Schedule jobEvent for execution after jobEvent.executionTime
% - Upon job completion, trigger deallocateResources for jobEvent

% Function deallocateResources(jobEvent):
% - Add jobEvent.requiredCores back to availableCores
% - If the priority queue is not empty:
% - Get the next jobEvent from the queue
% - Attempt to allocate resources for the next jobEvent

% Function mainSimulationLoop():
% - While simulation is running:
% - If new jobEvent arrives, call scheduleJobEvent(jobEvent)
% - Process any jobEvents whose execution time has elapsed
% - For each completed jobEvent, call deallocateResources(jobEvent)
% - Report using SST::Statistics
% \end{lstlisting}

\begin{algorithm}
\caption{Resource Allocation/Deallocation}
\label{alg:resource-allocation-deallocation}
\begin{algorithmic}[1]

\STATE \textbf{Initialize:}
\STATE \quad Define totalCores, availableCores
\STATE \quad Create an empty priority queue for JobEvents based on their priority

\STATE

\STATE \textbf{Function scheduleJobEvent(jobEvent):}
\STATE \quad Add jobEvent to the priority queue
\STATE \quad Attempt to allocate resources for jobEvent

\STATE

\STATE \textbf{Function allocateResources(jobEvent):}
\IF{availableCores $\geq$ jobEvent.requiredCores}
    \STATE Deduct jobEvent.requiredCores from availableCores
    \STATE Schedule jobEvent for execution after jobEvent.executionTime
    \STATE Upon job completion, trigger deallocateResources for jobEvent
\ENDIF

\STATE

\STATE \textbf{Function deallocateResources(jobEvent):}
\STATE \quad Add jobEvent.requiredCores back to availableCores
\IF{the priority queue is not empty}
    \STATE Get the next jobEvent from the queue
    \STATE Attempt to allocate resources for the next jobEvent
\ENDIF

\STATE

\STATE \textbf{Function mainSimulationLoop():}
\WHILE{simulation is running}
    \IF{new jobEvent arrives}
        \STATE call scheduleJobEvent(jobEvent)
    \ENDIF
    \STATE Process any jobEvents whose execution time has elapsed
    \FOR{each completed jobEvent}
        \STATE call deallocateResources(jobEvent)
    \ENDFOR
    \STATE Report using SST::Statistics
\ENDWHILE

\end{algorithmic}
\end{algorithm}

We chose SST due to number of its unique features. \textit{First}, SST provides parallel discrete event simulation capability to achieve performance.
% , efficiently managing the system's computational and memory resources. 
% This dynamic scheduling process is crucial for maintaining system performance, especially in the face of fluctuating job demands and resource availability. 
% The SST's event-driven architecture and cycle-accurate simulation capabilities plays a pivotal role in this process. SST's advanced synchronization mechanisms, including conservative synchronization strategies, ensure high-fidelity simulation of complex HPC systems~\cite{rodrigues2011structural}.
\textit{Second}, SST's architecture design is inherently modular. It incorporates various elements, such as \texttt{Merlin} for network modeling and \texttt{memHierarchy} for cache simulations. Each element serves specialized purposes within the simulation framework. 
% Each module serves specific purposes and contains many features. 
For example, \texttt{Merlin} facilitates the simulation of diverse network topologies such as dragonfly, torus, mesh, and fat-tree, alongside various routing algorithms to simulate data packet traversal through the network. It enables users to specify network parameters such as link bandwidth, latency, and routing policies. Meanwhile, \texttt{memHierarchy} is dedicated to simulating the memory subsystem of computer architectures and supports modeling different levels of the memory hierarchy, including caches, main memory, and storage. 
% It accommodates various cache coherence protocols, such as MESI, MSI, and MOESI, and permits the specification of cache sizes, associativity, and replacement policies. 
\textit{Third}, SST supports simulation of a wide range of applications -- from computational physics to data analytics -- and provides a comprehensive framework for accurate job execution simulation~\shortcite{rodrigues2011structural}. Our simulator leverages SST's robust event management capabilities~\cite{sst-elements} to model the job execution and captures the  dynamics of HPC job interactions. We can perform real-time monitoring of jobs via collecting SST's statistics.

% \subsubsection{Serialization of TaskEvent Class}

%\begin{lstlisting}[language=C++, caption={Serialization Method Implementation for %TaskEvent Class}, label=lst:task-event-serialization]
%TaskEvent(const std::string& id, int cores, double time, size_t memory) : 
%        Event(), jobID(id), requiredCores(cores), executionTime(time), requiredMemory(memory), state(QUEUED) {}

%    void serialize_order(SST::Core::Serialization::serializer& ser) override {
%        Event::serialize_order(ser);
%        ser & jobID;
%        ser & requiredCores;
%        ser & executionTime;
%        ser & requiredMemory;
%        ser & state; 
%    }
%    ImplementClassName("TaskEvent");
%\end{lstlisting}

% \subsection{Job Departure and Statistics Collection}

After job execution, our simulator collects and analyzes job statistics utilizing SST’s statistics framework. We gather statistics such as job durations, resource utilization, and system efficiency. The data collection and analysis helps in evaluating the effectiveness of different scheduling algorithms. The comprehensive statistics provided by SST enables us to refine our scheduling strategies continually.
% ensuring that our simulator remains at the forefront of HPC job scheduling and resource management research.

\section{Workflow Integration in SST Job Scheduling}

%\subsection{Introduction to Workflow Integration}

Current HPC systems increasingly demand simulations that not only manage isolated tasks, but also efficiently handle complex workflows consisting of multiple inter-dependent tasks. These workflows typically encapsulate scientific processes or business operations requiring robust scheduling and resource management. In this section, we introduce the workflow management component we incorporated within the SST. This component aims at simulating and managing the dynamics and dependencies of modern HPC workflows.
\note{The components supports both manually created and automatically generated directed acyclic graphs (DAG) to manage task dependencies. 
% This dual capability allows users to tailor the workflow complexity to their specific needs, facilitating both straightforward and complex scientific computing simulations. 
Other job scheduling simulators (e.g.,  CQsim, Alea 2) either do not support workflow dependencies or offer limited support, such as simple precedences or chaining of jobs. 
The proposed workflow integrates directly into the core simulation environment of SST and enhances its utility for dynamic HPC job behavior simulation.}

\subsection{Task Representation Design}

The \texttt{Task Scheduler} module models the individual computational jobs within a workflow. It encapsulates all the necessary information and behavior of these jobs. Following are some important attributes of this module.
\begin{itemize}
    \item \textbf{task\_id:} This is a unique identifier for the task.
    \item \textbf{execution\_time:} This represents the estimated time to execute the task. This could be based on computational complexity or historical data.
    \item \textbf{resource\_requirements:} This field represents required resources for the task, such as CPU cycles, memory size, and I/O bandwidth.
    \item \textbf{dependencies:} This represents a list of task IDs that must be completed before a particular task can start execution.
    \item \textbf{state:} This field represents the current state of the task (e.g., waiting, running, completed).
\end{itemize}

Within the task representation class, we check if all dependencies of the task have been satisfied. We also keep on checking if the state of a task has changed from waiting/running to completed. Once we detect that the state for a task is ``completed'', we trigger the rest of the tasks that have a dependency on the specific task that just completed. This helps in managing the execution order among the workflows in a particular job.

\subsection{Workflow Management Component}

The \texttt{Workflow Management} module in SST is designed to handle the details of task dependencies and resource allocation essential for scientific computing. Following are the key features of this module:

    % \textbf{Attributes:}
\begin{itemize}
    \item \textbf{tasks:} This field represents a dictionary or list holding all \texttt{Task} objects within the workflow.
    \item \textbf{workflow\_id:} This is a unique identifier for the workflow.
    \item \textbf{dependencies:} This is a representation of all dependencies in the workflow, potentially as a directed acyclic graph (DAG), where nodes are tasks and edges are dependencies.
\end{itemize}

Within the \texttt{Workflow Management} module, we process the tasks to identify and organize their dependencies, essential for scheduling and execution. We identify the dependencies of the tasks so we can schedule them to run upon the completion of the task(s) they are dependent on.
% once the task(s) their start is dependent upon completes running. 
We then return a list of tasks that are ready to be executed (i.e., all their dependencies have been satisfied).

\textbf{DAG Representation.} The DAG structure is a key modeling feature of workflows. It allows for efficient detection of ready tasks, and ensures that tasks are executed in the correct order. We implement the DAG using adjacency lists, due to the size and complexity of workflows~\shortcite{gupta2017generation}.
%~\cite{Baeldung2021}.

\textbf{Scheduling and Execution.} The workflow management involves a scheduler component that interacts with the \texttt{Workflow Management} module to allocate resources to Task objects based on their ready state and resource requirements. The \texttt{Workflow Management} module is responsible for triggering events in the SST simulation environment, corresponding to task starts and completions, to drive the simulation forward.
    % \textbf{Workflow Representation}: Workflows are modeled as directed acyclic graphs (DAGs), where each node represents a computational task, and edges represent dependencies. This structure ensures that SST can accurately simulate the execution order based on dependency resolutions.

    % \textbf{Task Representation}: Each task within a workflow is defined by its resource requirements (CPU cores, memory), execution time, and state (queued, running, completed). This granular representation allows SST to dynamically allocate resources and manage task states through its simulation lifecycle.

    % \textbf{Scheduling Algorithms}: Initially, we incorporated a First Come First Serve (FCFS) scheduling algorithm. This basic scheduling mechanism serves as a foundation for introducing more complex algorithms as the system evolves. FCFS treats tasks in the order they are submitted, which, while straightforward, sets the stage for more sophisticated scheduling techniques that will address the dynamic nature of workflow executions.

\begin{figure}[t!]
    \centering
    \includegraphics[width=0.55\textwidth]{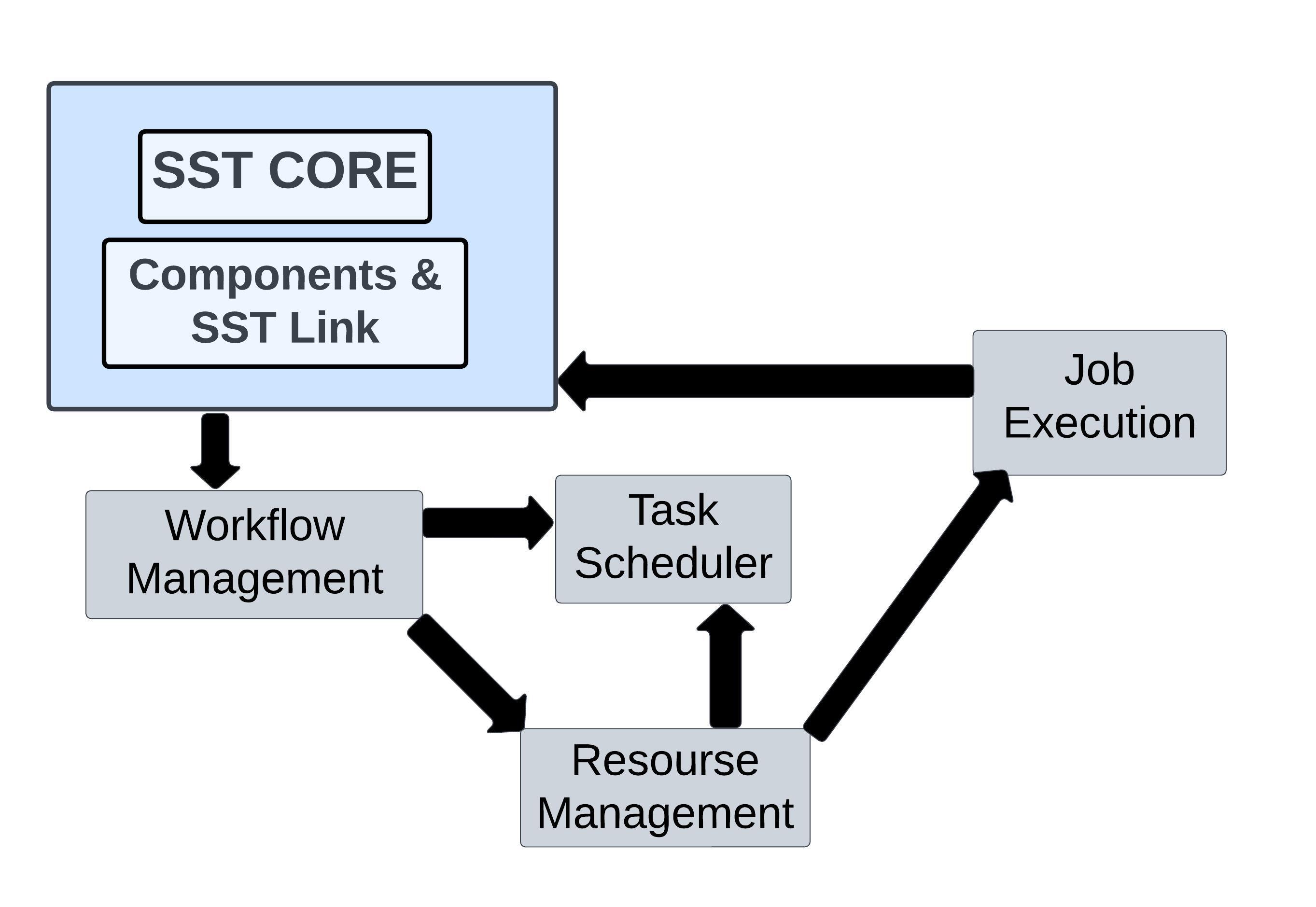}
    \caption{The workflow management components.}
    \label{fig:workflow}
\end{figure}

Figure~\ref{fig:workflow} illustrates the integration of workflow management within the SST. It shows various components (e.g., \texttt{SST Core}, \texttt{Workflow Management}, \texttt{Task Scheduler}, \texttt{Resource Management}, and \texttt{Job Execution}) that are interconnected to facilitate simulation of HPC workflows. The \texttt{SST Core} coordinates between the \texttt{Workflow Management} system that handles task scheduling based on dependencies and resources, and the \texttt{Resource Management} system that allocates necessary resources to execute tasks. \texttt{Job Execution} processes the actual computation tasks, feeding results back into the \texttt{SST Core}, completing the cycle of task management and execution.

% \subsection{Dynamic Scheduling and Resource Allocation}

% Future enhancements to SST’s workflow management component will focus on integrating advanced scheduling and resource allocation strategies. These enhancements include:

% \begin{itemize}
%     \item \noindent Dynamic Resource Scheduling: Unlike static scheduling, dynamic resource scheduling will allow SST to adjust resource allocations in real-time based on task priorities and current system load, improving overall system efficiency and reducing job wait times.
    
%     \item \noindent Preemption Capability: Implementing preemption will enable SST to interrupt lower priority tasks in favor of higher priority ones. This is crucial for workflows requiring immediate response to changing conditions or emergency scenarios~\cite{Preemption}.
% \end{itemize}

\subsection{Workflow Input Specification}

To specify workflows in our system, we use a JSON-based input format that defines tasks, their resource requirements, execution times, and dependencies. Below is an example of the JSON input format. The example illustrates how to define jobs, their dependencies, and the use of files as input and output within the workflow. This allows for efficient task management and resource utilization in a structured, JSON-based input format.
\vspace{\baselineskip}
\vspace{\baselineskip}

\begin{lstlisting}[language=json, caption=Workflow input format.]
{
  "tasks": [
    {"id": 1, "execution_time": 100, "resources": {"cpu": 2, "memory": 1024}, "dependencies": []},
    {"id": 2, "execution_time": 150, "resources": {"cpu": 1, "memory": 512}, "dependencies": [1]},
    {"id": 3, "execution_time": 200, "resources": {"cpu": 1, "memory": 512}, "dependencies": [1]},
    {"id": 4, "execution_time": 300, "resources": {"cpu": 2, "memory": 1024}, "dependencies": [2, 3]}
  ],
  "resources_available": {"cpu": 10, "memory": 8192},
  "scheduling_policy": "Static",
  "preemption": false
}
\end{lstlisting}

% The workflow management component's preliminary validation involved testing with simplified workflow scenarios inspired by well-established workflow systems such as Pegasus and Apache Airflow~\cite{ExpOfWMS}. These tests were crucial for ensuring that SST could handle basic workflows effectively while adhering to dependency constraints and maintaining performance integrity. The results from these initial tests have been promising, demonstrating SST's capability to manage workflows with varying complexities and dependencies accurately.

For validation (Section~\ref{sec:results}), we tested with simplified workflow scenarios such as generated by well-established workflow systems (e.g., Pegasus and Apache Airflow~\shortcite{mitchell2019exploration}). By using these tests we were able to ensure that SST could handle basic workflows effectively, while adhering to dependency constraints and maintaining performance integrity. 
% The results from these initial tests have been promising, demonstrating SST's capability to manage workflows with varying complexities and dependencies accurately.

\note{Our simulator, leveraging the capabilities of the SST, creates the DAGs for managing task dependencies in HPC workflows. Users have the flexibility to manually define DAGs, which is crucial for workflows with well-understood and stable task relationships. This manual method ensures that the execution order adheres strictly to the user's specifications and control.
% In comparison, simulators like CQsim primarily focus on static job scheduling without inherent support for dynamic task dependencies typical in complex workflows. Other tools, such as Alea 2 and the Slurm Simulator, offer varying degrees of workflow management but do not typically provide the same level of integration between DAG management and real-time adaptability as facilitated by SST’s modular and extensible framework.
}

\section{Experiments}
In this section, we present the validation and scalability evaluation of our proposed job scheduler simulator.

\subsection{Experiment Setup}
We leverage job trace from the Grid Workloads Archive (GWA)~\shortcite{iosup2008grid}. We validated our simulation output against CQsim~\cite{CQsim}, an event-based cluster scheduling simulator. Specifically, we use the GWA-DAS2 trace from GWA to simulate real-world HPC environments. The GWA-DAS2 trace contains about 1,124,772 jobs. Each job includes attributes such as job submission time, start/end time, memory usage, run time, wait time of jobs, the user information, the project or group associated, and the outcome of the job execution. The GWA-DAS2 trace provides a diverse range of job submissions, sizes, and resource requirements, offering a comprehensive basis for evaluating the scheduling algorithms. We also used the SDSC-SP2 log~\cite{sdsc_sp2_log} from the Parallel Workloads Archive~\cite{parallel_workloads_archive} to test the scalability of our job scheduler simulator. The SDSC-SP2 log contains attributes of jobs, such as job number, job submission time, wait time, run time, number of allocated processors, average CPU time used, requested time, requested memory, and queue number~\cite{sdsc_sp2_log}. The job log contains 73,496 job information. 
\note{We use the trace data to determine the start time and end time of jobs in our simulator.}
% \note{In our simulator, we use fixed execution times for each job. These times are set during the simulation setup, based on initial resource requests and estimated job durations. This approach guarantees consistent and repeatable results, essential for analyzing the impact of different scheduling algorithms and resource management strategies. Given our focus on assessing the theoretical limits and efficiency of scheduling policies in a controlled environment, the fixed-time approach offers the necessary rigor and simplicity.} 
\note{We also utilize job trace data from various epigenomic sequencing projects to simulate real-world computational environments in bioinformatics. Specifically, our analysis includes the processing of sequences such as 4seq, 5seq, and 6seq, which are indicative of different epigenomic data complexities~\shortcite{juve2013characterizing}.} 

CQSim is a Python-based discrete-event-driven cluster scheduling simulator that evolved from QSim to enhance functionalities. It is designed with a modular Python architecture for reusability, extensibility, and efficiency. It simulates job submissions, allocations, and executions based on workload traces~\cite{CQsim}. CQSim has been used in diverse HPC job scheduling mechanism evaluation such as energy-aware scheduling~\shortcite{yang2013integrating}, fault-aware scheduling~\shortcite{tang2009fault}, etc.

In addition to leveraging job traces from the Grid Workloads Archive (GWA) and validating against CQsim, we plan to extend our testing to encompass real-life workflow simulations using well-known workflow management systems like Pegasus. Pegasus is a robust workflow management system designed to automate the execution of complex workflows over diverse computing environments, including clouds and grids. It transforms high-level abstract workflow descriptions into executable workflows mapped onto physical resources. Notably, Pegasus optimizes the execution of these workflows by managing data placement and task clustering, which minimizes data transfer and enhances overall computational efficiency.
Specifically, integrating real-life workflows from Pegasus into the Structural Simulation Toolkit (SST) environment will allow us to explore and evaluate the scheduler's effectiveness under realistic scientific computational loads. These workflows, often used in disciplines such as bioinformatics, astronomy, and climate science, provide a rigorous testbed for assessing SST's capabilities in handling complex dependency-driven job execution. The integration of Pegasus workflows will allow us to simulate more realistic and practical scenarios, ensuring that SST can support advanced scheduling techniques effectively in a high-performance computing (HPC) context. This approach will not only validate SST's performance under realistic conditions but also showcase its potential to adapt and manage workflows commonly used in scientific research, thereby enhancing its applicability and robustness in real-world HPC scenarios.

\subsection{Results}
\label{sec:results}

% Using  These graphs are a validation \& illustration of the comparative analysis of the Task Scheduler component \& the scheduling algorithms:

    \begin{figure}[h!]
    \centering
    \subfigure[Number of occupied nodes over simulation time]{
        \label{fig:nodes_occupied}
        {\includegraphics[width=0.43\textwidth]{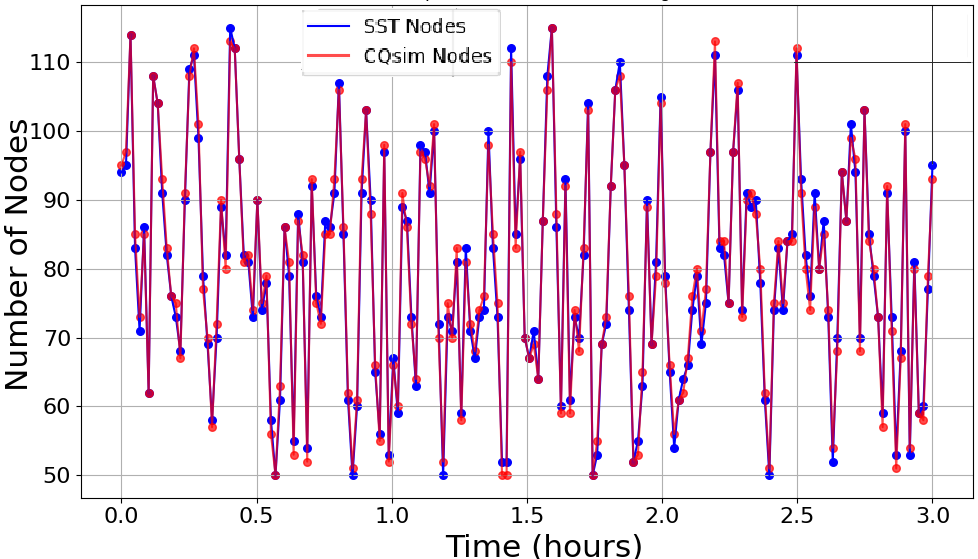}}
    }
    \subfigure[Number of jobs over simulation time]{
        \label{fig:jobs_running}
        {\includegraphics[width=0.49\textwidth]{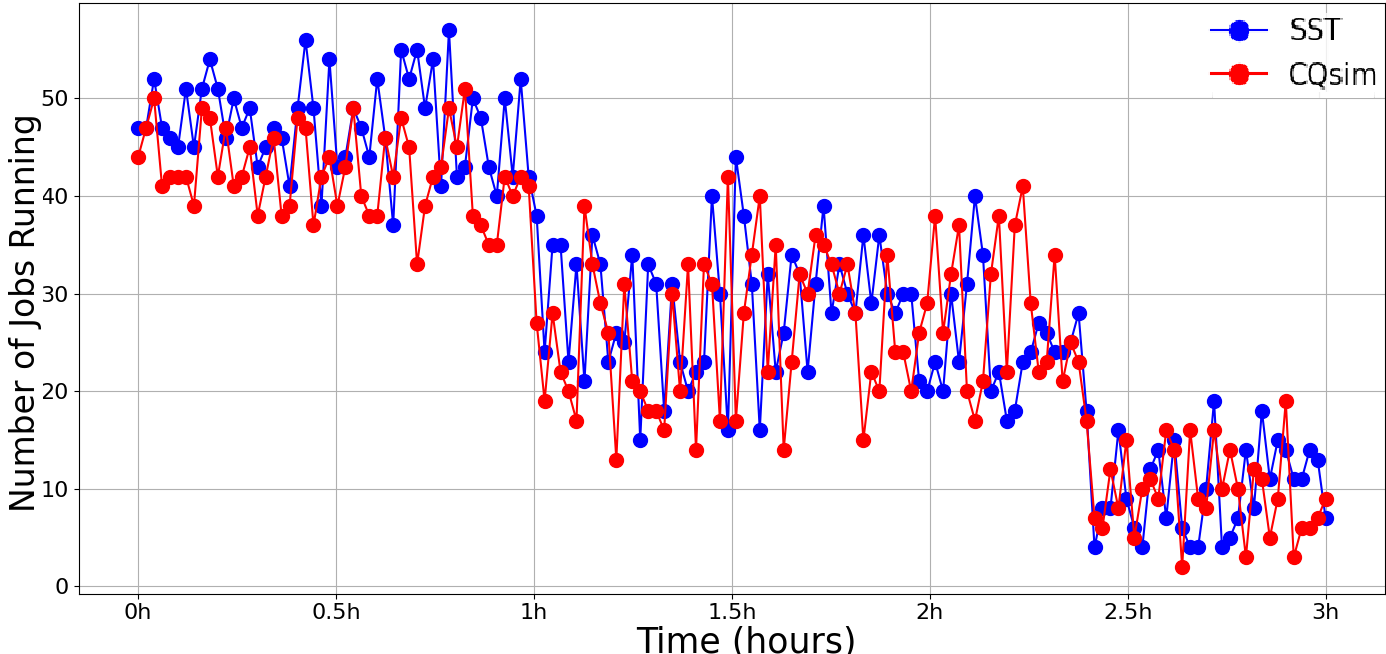}}
    }
    \caption{Comparison of our simulation output with CQSim.}
    \label{fig:sst_workflow}
    \end{figure}
In this section, we present the results based on our simulator experiments.  
Figure~\ref{fig:nodes_occupied} presents the occupancy of nodes throughout the simulation duration. We use the DAS-2 log for this simulation. We compare our simulation output with the CQsim simulation. As can be seen in the figure, the number of nodes occupied using our simulator is similar to the  CQsim simulation output. 
% This consistency underscores the algorithm's robustness in diverse simulation environments.

Figure~\ref{fig:jobs_running} outlines the temporal distribution of job executions during the simulation. We observe a gradual decline in the number of active jobs. This trend is consistent across both simulation platforms, reinforcing the impact of job size on system throughput. 
The distribution of job sizes within the trace log (generally categorized as small, medium, and large across the initial, middle, and final phases, respectively) facilitates a comprehensive evaluation of the scheduling algorithm's performance across a range of workload scenarios.

% The observed data underscores the effectiveness of our scheduling approach in managing workloads of varying sizes \& complexities, thereby validating its applicability for efficient resource allocation in high-performance computing environments.

\begin{figure}[t!]
	\centering
	\subfigure[Wait time validation against CQSim.]{
		\label{fig:wait_cqsim}
		{\includegraphics[width=0.46\linewidth]{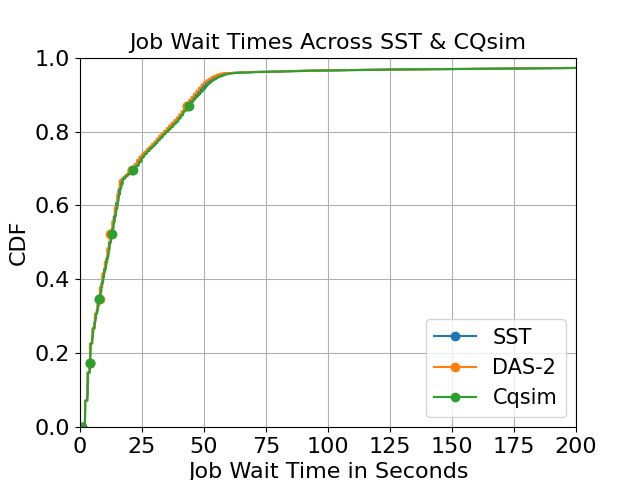}}	}
	\subfigure[Wait time of different job scheduling algorithms]{
		\label{fig:wait_scheduling}
		{\includegraphics[width=0.46\linewidth]{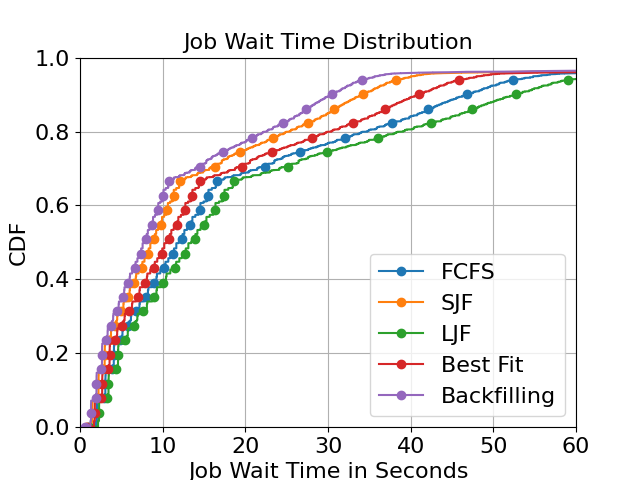}}	}
	\caption{Job wait time validation and comparison of different scheduling algorithms.}
	\label{fig:job_wait_time}
\end{figure}

Figure~\ref{fig:job_wait_time} presents wait time analysis of our simulator. In Figure~\ref{fig:wait_cqsim}, we compare wait time measurements against CQSim. As can be seen in the figure, the job wait time of our simulator closely matches with the measurements from both CQSim and DAS-2 job trace. This validates our simulator's accuracy in prediction. In Figure~\ref{fig:wait_scheduling}, we present a comparative analysis of the five job scheduling algorithms. Although \texttt{FCFS} provides simplicity (ideal for jobs of uniform size) it has lower resource utilization. \texttt{FCFS} with \texttt{Best Fit} improves resource matching but does not significantly improve job completion times. \texttt{FCFS} with \texttt{Backfilling} maximizes resource utilization by intelligently filling scheduling gaps, while \texttt{SJF} reduces average job completion times by focusing on shorter jobs. In contrast, \texttt{LJF} is less efficient, often resulting in poorer utilization due to its preference for longer jobs. 

\begin{figure}[t!]
	\centering
	\subfigure[Running DAS-2 on SST across multiple ranks]{
		\label{fig:sst_scalability}
		{\includegraphics[width=0.47\linewidth]{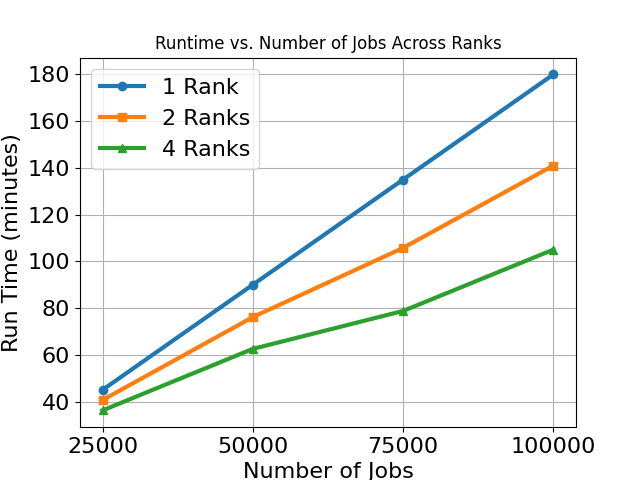}}	}
	\subfigure[Running SDSC-SP2 on SST across multiple ranks]{
		\label{fig:sdsc_scalability}
		{\includegraphics[width=0.47\linewidth]{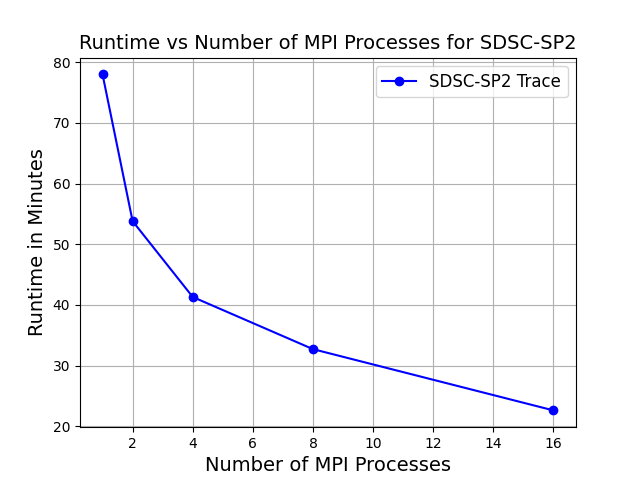}}	}
	\caption{Scalability of our proposed simulator.}
	\label{fig:scalability}
\end{figure}

\begin{figure}[]
	\centering
	% \subfigure[Workflow scalability.]{
		{\includegraphics[width=0.51\linewidth]{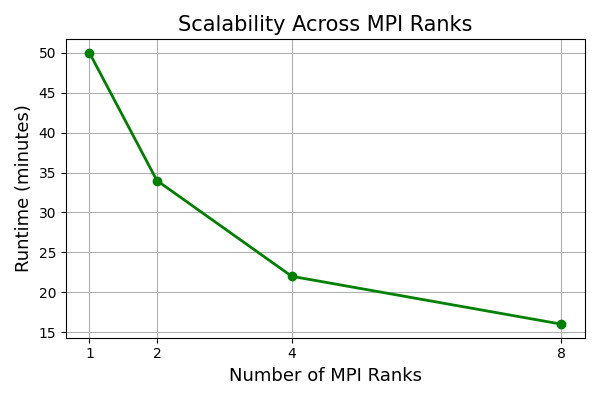}}	
  \caption{Scalability of workflow simulation.}
  \label{fig:workflow_scalability}
  % }
\end{figure}

Figure~\ref{fig:scalability} demonstrates the parallel performance of our job scheduler when executing the DAS-2 \& SDSC-SP2 workload on SST. Figure~\ref{fig:sst_scalability} shows performance for the DAS-2 log, while Figure~\ref{fig:sdsc_scalability} presents result for the SDSC-SP2 trace. As can be seen in the figures, simulator performance scales well as the number of MPI ranks increases. 
% This trend confirms the scheduler's capacity to scale efficiently hence leading to improved throughput \& resource utilization. The scalability of the scheduler is validated using the GWA-DAS2 \& SDSC-SP2 workload, offering a comparative view against different job sets \& computational demands to further substantiate the flexibility \& effectiveness of our scheduler. 
We also notice from Figure~\ref{fig:sst_scalability} that as the job sizes increased, we achieve greater speedup. \note{SST's scalability property (e.g., support of parallel discrete-event simulation) and our model's optimizations enable good scalability.} Figure~\ref{fig:workflow_scalability} shows scalability result of the proposed workflow-based HPC job simulation.
% Unlike CQSim, which struggles with scalability due to its less dynamic architecture, our scheduler ensures that performance gains are realized even as system complexity grows.

Figure~\ref{fig:workflow_scalability} demonstrates the parallel performance the workflow based scheduler while running the Galactic workflow from Pegasus workflow gallery. The Galactic Plane workflow uses the Montage image mosaic engine to transform all the images in 17 sky surveys to a common pixel scale of 1 second or arc, where all the pixels are co-registered on the sky and represented in Galactic coordinates and the Cartesian projection ~\cite{PegasusGallery2024}. As can be seen in the figures, simulator performance scales well as the number of MPI ranks increases.

% Figure~\ref{fig:epigenomics_cdf} shows the Cumulative Distribution Function (CDF) of average CPU utilization for different epigenomic sequencing datasets (4seq, 5seq, 6seq). This graph is used to illustrate the efficiency of computational resources across various epigenomic sequencing workflows. It helps validate our computational processes, indicating their capability to handle diverse epigenomic data.

% \begin{figure}[ht]
%     \centering
%     \includegraphics[width=0.7\linewidth]{Figures/Epigenomics_CDF.png}
%     \caption{Average CPU Usage Validation}
%     \label{fig:epigenomics_cdf}
% \end{figure}

\begin{figure}[]
    \centering
    \includegraphics[width=0.51\linewidth]{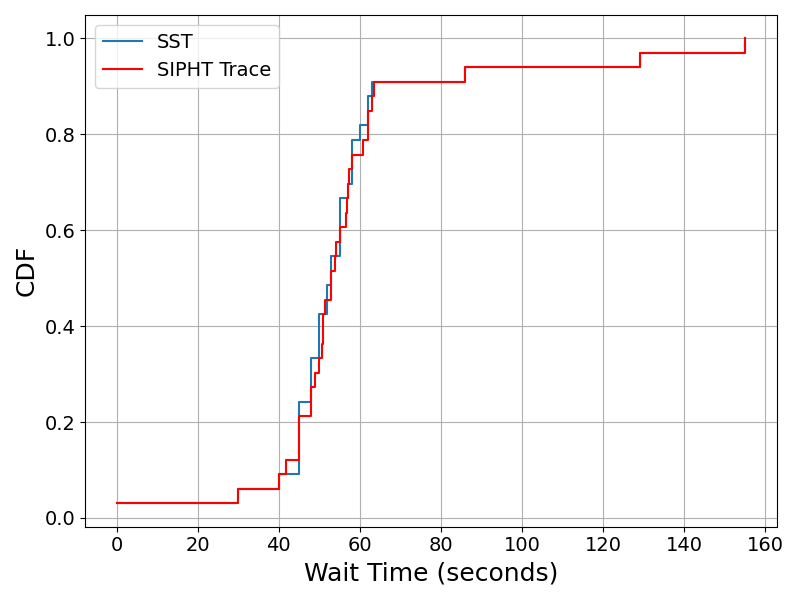}
    \caption{Job wait time validation of workflow simulation.}
    \label{fig:sipht_wait_time_cdf}
\end{figure}

\note{Figure~\ref{fig:sipht_wait_time_cdf} presents the wait time comparison for the SIPHT workflow and our workflow simulator. The SIPHT workflow is an application in bioinformatics for automating the search for untranslated RNAs (sRNAs) within bacterial replicons~\cite{SIPHT}. As can be seen in Figure~\ref{fig:sipht_wait_time_cdf}, the workflow wait time of our simulator closely matches with real-life measurements of the SIPHT workflow, validating our workflow component of the simulator.
% The performance, as demonstrated by the wait time distribution, confirms our simulator's capability to handle workflow dependencies \& manage job scheduling efficiently in real-world applications with workflows.
}

\section{Conclusion}
In this paper, we propose a job scheduling simulator for scheduling HPC jobs on large-scale HPC platform. Our proposed approach features integration of job scheduling and resource management components on event-driven simulator, SST. We validate our proposed model against job scheduling simulator using HPC job traces. We demonstrate accuracy and parallel performance of our simulator in HPC job scheduling and resource management.   
\note{Our workflow component  demonstrates the potential to simulate HPC job behaviors with workflow dependencies. The preliminary results confirm its effectiveness in handling task dependencies and resource allocations.}
% \note{, with further enhancements like automatically generating DAGs to handle the dynamic nature of scientific computing applications planned to expand its capabilities. This  development approach ensures that as the needs of HPC environments evolve, our simulator will continue to provide relevant and impactful solutions.}

% In the next step, we will include better resource management techniques for architecture designers to explore and identify optimal resources. We will further aim to explore advanced machine learning techniques (e.g., RL) to improve the scheduling performance. 

In the next step, we plan to carefully study HPC job managers (e.g., SLURM) and develop models to support heterogeneous job and multi-cluster operation. We also plan to study other new job scheduling strategies (e.g., using reinforcement techniques) to improve job scheduling performance. We will incorporate these advanced job scheduling techniques in our simulator. 
% We will further explore job scheduling and resource management technique cloud continuum that includes HPC and cloud jobs.  
% Looking forward, we aim to explore the integration of reinforcement learning techniques to enhance scheduling accuracy and system adaptability, particularly in anticipation of the exascale computing era. This will include comparative performance analyses against established simulators like CQSim, focusing on scalability and adaptability to evolving HPC architectures.
We also aim to enhance SST's workflow management by integrating dynamic scheduling and preemption capabilities to better adapt to fluctuating workloads and resource availability. We will also develop support for complex, interdependent workflows, drawing on features from established systems like Pegasus and Apache Airflow. These improvements will enable more efficient handling of diverse scientific workflows, aligning SST more closely with modern high-performance computing needs.
In the future, we  will also focus on integrating advanced resource scheduling (e.g., dynamic scheduling) and resource allocation strategies (e.g., preemption capability) to SST’s workflow management component. 
% For example, unlike static scheduling, dynamic resource scheduling will allow SST to adjust resource allocations in real-time based on task priorities and current system load, improving overall system efficiency and reducing job wait times. Furthermore, implementing preemption capability will enable SST to interrupt lower priority tasks in favor of higher priority ones. 
% This is crucial for workflows requiring immediate response to changing conditions or emergency scenarios.

\section*{ACKNOWLEDGMENTS}
% This work was supported in part by the National Science Foundation EPSCoR Program under NSF Award \# OIA-1655740 and an  ASPIRE grant  from  the  Office  of  the  Vice  President for Research at the University of South Carolina.
This work is supported in part by the U.S. National Science Foundation under grants CNS-2300124, and OAC-2411456, and supported by the U.S. Department of Energy, Office of Science under Contract No. DE-AC02-06CH11357.

% Reducing font size (to 9pt) for References & Author Biagraphies
\footnotesize

% Please don't exchange the bibliographystyle style
\bibliographystyle{wsc}

% AUTHOR: Include your bib file here
\bibliography{references}

\section*{AUTHOR BIOGRAPHIES}

\noindent {\bf \MakeUppercase{ABUBEKER ABDURAHMAN}} is currently an undergraduate Research Assistant at The University of Toledo where he is pursuing a Bachelor's of Science in Computer Science \& Engineering. His email address is \email{abubeker.abdurahman@rockets.utoledo.edu}. \\

\noindent {\bf \MakeUppercase{ABRAR HOSSAIN}} is currently working as a Graduate Research Assistant at The University of Toledo where he is also pursuing a Master's in Computer Science. His research focuses on High-performance computing (HPC) and Stochastic Modeling, Control, and Optimization. His email address is \email{abrar.hossain@rockets.utoledo.edu}. \\

\noindent {\bf KEVIN A. BROWN} is the inaugural Walter Massey Fellow at Argonne where he conducts research in the Mathematics and Computer Science (MCS) Division. He received his M.Sc. and Ph.D. in Mathematical and Computing Sciences at the Tokyo Institute of Technology in 2014 and 2018, respectively. He worked as a Postdoctoral Appointee in the Argonne Leadership Computing Facility (ALCF) where he optimized network designs and communication strategies for Exascale supercomputers by utilizing advanced routing, quality-of-service, and congestion management mechanisms. His email address is \email{kabrown@anl.gov}.\\

\noindent {\bf KAZUTOMO YOSHII} is a Software Development Specialist at Argonne National Laboraotry. He received a M.S. in computer science from Toyohashi University of Technology in Japan. His research interests include hardware/software co-design, heterogeneous and reconfigurable computing/field-programmable gate arrays, data movement and memory management, and edge computing architecture. His email address is \email{kazutomo@mcs.anl.gov}.\\

\noindent {\bf \MakeUppercase{KISHWAR AHMED}} is an Assistant Professor in the Department of Electrical Engineering and Computer Science at the University of Toledo. He received a Ph.D. degree from School of Computing and Information Sciences, Florida International University. His research interests include high-performance computing, energy-efficiency, and optimization. His email address is \email{kishwar.ahmed@utoledo.edu}.

\end{document}